\documentclass[11pt]{article}

\def\ga{\mathrel{\raise.3ex\hbox{$>$\kern-.75em\lower1ex\hbox{$\sim$}}}}
\def\la{\mathrel{\raise.3ex\hbox{$<$\kern-.75em\lower1ex\hbox{$\sim$}}}}

\def\I_M{{I_{\scriptscriptstyle M\times M}}}

\begin{document}

\thispagestyle{empty}
\rightline{IP/BBSR/2002-10}
\rightline{{\tt hep-th/0205180}}

\vskip 2cm \centerline{ \Large \bf Bouncing and cyclic universes from 
brane models}

\vskip .2cm

\vskip 1.2cm

\centerline{ \bf Sudipta Mukherji $^{a,b}$ and Marco Peloso $^b$\footnote{New Address: C.I.T.A., University of Toronto, 60 St. George Street, Toronto, Ontario, Canada M5S 3H8}}
\vskip 10mm \centerline{ \it $^a$ Institute of Physics, 
Bhubaneswar-751 005, India} 
\vskip 3mm 
\centerline{\it $^b$ Physikalisches Institut, Universit\"at 
Bonn,} 
\centerline{\it Nussallee 12, D-53115 Bonn, Germany} 
\vskip 1.2cm

\centerline{\tt  mukherji@iopb.res.in, peloso@cita.utoronto.ca 
}

\vskip 1.2cm

\begin{quote}

We consider a D3-brane as boundary of a five dimensional charged anti de
Sitter black hole. We show that the charge of the black hole induces a
regular cosmological evolution for the scale factor of the brane, with a
smooth transition between a contracting and an eventual expanding phase.
Simple analytical solutions can be obtained in the case of a vanishing
effective cosmological constant on the brane. A nonvanishing cosmological
constant, or the inclusion of radiation on the brane, does not spoil the
regularity of these solutions at small radii, and observational
constraints such as the ones from primordial nucleosynthesis can be easily
met. Fluctuations of brane fields remain in the linear regime provided the
minimal size of the scale factor is sufficiently large. We conclude with
an analysis of the Cardy-Verlinde formula in this set up.

\end{quote}

\newpage
\setcounter{footnote}{0}

\section{Introduction}

Motivated by string/M theory, the AdS/CFT correspondence, and the
hierarchy problem of particle  physics, brane-world models were studied
actively in recent  years \cite{HW}-\cite{RSt}. In these models, our
universe is realised as a  boundary of a higher  dimensional space-time
(the so called bulk). In particular, a well studied example is when the
bulk is an AdS space. The gravitational interaction among matter on this
brane is found to be described by standard laws when one considers
distance scales much larger than the inverse of the AdS mass
scale~\cite{gata}.

In the cosmological context, embedding of a four dimensional 
Friedmann-Robertson-Walker (FRW) universe was also considered when the
bulk is described by AdS or AdS black hole, see for example
\cite{TN}-\cite{NOO}. In the latter case, the mass of the black hole was
found to effectively act as an (invisible) energy density on the brane
with the same equation of state of radiation. In another line of
interesting development, initiated in \cite{EV,SV}, holographic principle
was studied in the FRW universe filled with such dark radiation.
Representing radiation as conformal matter and exploiting AdS/CFT
correspondence, a Cardy like formula~\cite{CARDY} for the entropy was
found for the universe. This is often referred to as Cardy-Verlinde
formula in recent literature. Furthermore, in \cite{EV}, a cosmological
entropy bound was proposed which unifies the Bekenstein bound for a
limited self-gravity system and the Hubble bound for a strong-gravity
system in an elegant way (for related works, see e.g.~\cite{NO,CMO,AP}).

In the present paper, we study the cosmology of a four  dimensional brane 
which constitutes the boundary of a charged AdS black hole background. By
introducing this extra charge parameter on the bulk, we find a host of 
interesting cosmological solutions for our brane universe. Their most
remarkable feature is that the charge term allows  for a {\it nonsingular}
transition between a contracting phase of the scale  factor of the brane,
and a following expanding stage.~\footnote{A qualitatively similar
behavior is shown by some solutions of the system described
in~\cite{BCG}.} The cosmological evolution of this model is thus free from
singularities. The latter reappear only when the bulk charge is taken to
zero. 

In the next section of the paper, we briefly review the charged AdS black
hole background. In section ~\ref{crit} we then set up the Hubble equation
for the scale factor of the brane, and analyse the induced four
dimensional cosmological evolution. We present exact solutions in the case
of a vanishing effective cosmological constant on the brane, which can be
achieved through a fine tuning of the brane tension and the five
dimensional cosmological constant. In the case of a flat or an open
geometry, the evolution is characterized by a nonsingular and smooth
transition between a contracting and an eventual expanding phase. For a
closed geometry, also the expanding phase turns into contraction at later
times, so that the whole evolution is cyclic. In section~\ref{radio} we
discuss the more physically relevant case in which radiation is present on
the brane. We discuss under which conditions the system can describe a
universe with a bounce followed by a radiation dominated phase compatible
with the constraints coming from the successful predictions of (standard)
primordial nucleosynthesis. In section ~\ref{pert} we discuss the
cosmological evolution of fluctuations of fields on the brane. We show
that they remain in the linear regime provided the minimum size of the
scale factor is sufficiently large. We also comment on the more general
issue of cosmological perturbations in this model. In section~\ref{CVF} we
briefly analyse the Cardy-Verlinde formula in the light of the explicit
solutions that we obtained. Finally, we present our conclusions.

\section{The five dimensional background\label{EPS}}

In this work we will consider a $3+1$ dimensional brane in a space-time
described by a $5$ dimensional charged AdS black hole. The background
metric is thus given by \cite{CEJM,CS}
\begin{equation}
ds_{5}^2 = - h(a) dt^2 + {{da^2}\over{h(a)}} + a^2 \gamma_{ij} dx^i dx^j,
\label{metric}
\end{equation}
where 
\begin{equation}
h(a) = k - {\omega_{4} M\over{a^{2}}} + {3 \omega_{4}^2 Q^2
\over {16 a^{4}}} + {a^2\over{L^2}}.
\label{component}
\end{equation}
Here, $k = 0 ,\, \pm 1$, corresponding to flat, open, or close geometries
of four dimensional subspaces at any given $a\,$. $M$ and $Q$ are,
respectively, the  Arnowitt-Deser-Misner mass and charge, and $L$ is the
curvature radius of space. $\gamma_{ij}$ is the metric  for a constant
curvature manifold $M^3$ with ${\rm Vol} (M^3) = \int d^3x {\sqrt
\gamma}$. ${\bf G_5}$ is the five dimensional Newton's constant and
$\omega_4 = 16 \pi {\bf G_5}/ 3  {\rm Vol} (M^3)$. It can be easily shown
that the above metric~(\ref{metric}) describes a  charged black hole with
two horizons only provided that
\begin{equation}
L^2 > \frac{3 \, x_H^2}{\omega_4 \, M - 2 \, k \, x_H} \,\,,
\label{boundl}
\end{equation}
where
\begin{equation}
x_H = \omega_4 \, M \left( \sqrt{1+ \frac{9 \, Q^2}{16 \, M^2}} - 1 \right)
\;,\;\; x_H = \frac{9\,\omega_4\,Q^2}{32\,M} \;,\;\; x_H = \omega_4 \, M \left( 1 - \sqrt{1 - \frac{9 \, Q^2}{16 \, M^2}} \right) \,\,,
\end{equation}
in the three cases $k=-1,\,0,\,1\,$, respectively. It can be also shown
that, when the limit~(\ref{boundl}) is satisfied, the quantity $a_H \equiv
\sqrt{x_H}$ represents a lower bound on the position of the larger of the
two horizons.

The electrostatic potential difference between the horizon and infinity is
given by the quantity ${\phi}$ which, following \cite{CEJM}, we choose to be 
\begin{equation}
{ \phi} = \Big({3\over{8}}\Big){\omega_{4} Q\over{a^{2}}}.
\label{potential}
\end{equation}
The entropy and the temperature associated with (\ref{metric}) are given by
\begin{eqnarray}
{\cal {S}} &=& {a_H^3 {\rm Vol}(M^3)\over{4 {\bf G}_5}},\nonumber\\ 
{\cal {T}}  &=& {h^\prime (a)\over{4\pi}}|_{a=a_H} = 
{4 a_H^2 + 2 k L^2 \over{4\pi L^2 a_H}} - {3 \omega_4^2 Q^2 
\over {32 \pi a_H^5}}.
\label{st}
\end{eqnarray}

Let us now consider a $(3+1)$-dimensional brane with a constant tension in
the background of (\ref{metric}). As discussed in \cite{SV}, if we regard
the brane as a boundary of the background AdS geometry, its location and
its induced metric become time dependent. More specifically, the induced
four dimensional metric describes a FRW Universe, with a scale factor
evolution determined both by the five dimensional geometry and by the
constituents of the brane itself. In the next two sections we discuss in
details the cosmological evolution of the four dimensional metric for the
geometry~(\ref{metric}). Here we note that the boundary CFT can be
determined only up to a conformal factor \cite{GKP,W}. Making use of this
fact, we rescale the boundary metric in the following form
\begin{equation}
ds^2_{CFT} = {\rm lim}_{a \rightarrow \infty} \Big[{L^2\over{a^2}} 
ds_5^2 \Big] = -dt^2 + L^2 d\Omega_4^2.
\label{bdary}
\end{equation}
As the thermodynamical quantities of the CFT at high temperature can be 
identified with those with the bulk adS black hole \cite{witten},
we get the following thermodynamical variables on the boundary CFT:
\begin{equation}
E = {L M\over a}, ~~\Phi = {L\phi\over a}, ~~T = {L{\cal{T}}\over a}.
\label{branest}
\end{equation}
These extra factors of $L/a$ appear following the scaling in 
(\ref{bdary}). The entropy $S$ of the CFT remains the same as 
(\ref{st}).

\section{Cosmological evolution of the four dimensional brane}~\label{crit}

Consider now a brane with tension $\Lambda_{\rm br}$ in the
background~(\ref{metric}). It can be shown (see~\cite{BGG,MV,CEG,SV,GP} for
details) that an observer on the brane experiences a  four dimensional FRW
universe
\begin{equation}
ds_{4}^2 = -d\tau^2 + a(\tau)^2 \gamma_{ij} dx^i dx^j,
\end{equation}
with a Hubble law given by
\begin{equation}
H^2 \equiv \left( \frac{\dot{a}}{a} \right)^2 = - \frac{k}{a^2} + {\omega_{4} M\over {a^{4}}} - 
{ 3 \omega_{4}^2 Q^2\over{16 a^{6}}}+ \frac{\Lambda_4}{3} \,,
\label{freedman}
\end{equation}
where $k=+1,-1,0$ correspond to a closed, flat or open geometry,
respectively, dot denotes derivative with respect to the physical time
$\tau$ on the brane, and where the effective four dimensional cosmological
constant $\Lambda_4 = \Lambda_{br} - 3/L^2\,$ gets a contribution from the
bulk cosmological constant and the brane tension $\Lambda_{br}\,$. By
tuning these two contributions, $\Lambda_4$ can be set to zero, in which
case the brane is denoted as {\it critical brane}. In the present section
we will discuss the cosmological evolution of an empty and critical brane, while the more general cases of a {\it non critical} brane and of a brane filled with radiation will be discussed in the next section.

By looking to the scaling with $a$ of the different terms  of
eq.~(\ref{freedman}), we see that the terms proportional  to $M$ and $Q$
behave, respectively, like the energy density of  radiation and of ``stiff
matter'' (i.e. with dominance  of the kinetic energy) on the brane.  In
the second case, the sign is  however opposite with respect to the
standard situation. Since this last  term dominates for sufficiently small
values of $a\,$, one may expect  that this sign difference could have
interesting cosmological  consequences. Indeed, we will see that it is
crucial in allowing  a nonsingular transition between a contracting and
an  expanding evolution of the scale factor $a\,$. 

The evolution of the system can be solved exactly, as one can most simply
realize by using conformal time  $\eta\,$, defined as $d\tau = a(\eta)
d\eta\,$. Let us first consider a closed four dimensional  geometry. In
this case case the solution is
\begin{equation}
a \left( \eta \right) = \sqrt{\omega_4 M\over 2} \Big[ 1 - c_1 ~\cos \left( 2 \, \eta \right) \Big]^{1\over 2},
~~~{\rm with}~~ c_1 = \sqrt{1 - {3 Q^2\over{4 M^2}}}.
\label{konesoln}
\end{equation}
The reality of $c_1$ (i.e. $Q < 2 M/{\sqrt 3}$) is a necessary condition
for the existence of a horizon in the five dimensional geometry, see
eq.~(\ref{boundl}), which we always assume to be the case. Hence, the
universe evolves periodically, with a four dimensional radius oscillating
between a maximal and a minimal size given by
\begin{equation}
a_{\rm max,min} = 
\sqrt{\frac{\omega_4 \, M}{2}} \, \Big( 1 \pm c_1 \Big)^{1/2} \,\,.
\label{minmax}
\end{equation}
Notice that we have used the freedom in setting the origin of conformal
time so that the minimal radius is reached at $\eta = n \, \pi\,$, where
$n$ is an integer number. From the bulk perspective, the brane  starts out
from the charged black hole at a distance $a_{\rm min}$ from the 
singularity  and moves away up to $a_{\rm max}$ as it  expands. At later
time, it collapses again at $a = a_{\rm min}$.

It is instructive to compare the case at hand with the situation in which
the brane is in the background of an AdS-Schwarzchild black hole. Also in
the latter situation, the cosmological evolution is governed by
eq.~(\ref{freedman}), but with with $Q=0$. This, in turn, means that the
time dependence of the scale factor is given by (\ref{konesoln}), with 
$c_1 = 1$. In this case, the scale factor starts from zero size 
and expands up to $a_{\rm max}$ before collapsing again to zero
size. From the bulk point of view, the brane originates from
the black hole singularity and at a later stage it collapses again into the
singularity. We, therefore, see that the effect of the background charge
$Q$ is rather non-trivial, since it makes the cosmology of the model free
from singularities.

For an open universe, we find the solution
\begin{equation}
a \left( \eta \right) = \sqrt{\omega M\over 2} \, \Big[ c_2 ~{\rm cosh } \left( 2 \, \eta \right)  -1 \Big]^{1\over 2}, ~~~~{\rm with}~~ c_2 
= \sqrt{1 + {3 Q^2 \over{4 M^2}}} \,.
\label{oscale}
\end{equation}
In this case, the brane is initially contracting, and then bounces to an
expanding phase. Again, we have set $\eta=0$ at the bounce. The minimal
radius is given by
\begin{equation}
a_{\rm min} = \sqrt{\frac{\omega_4 \, M}{2}} \, \Big( c_2 - 1 \Big)^{1/2} \,\,.
\end{equation}
As before, $a_{\rm min} \rightarrow 0\,$ as $Q \rightarrow 0\,$.

Finally, in the case of a flat universe we find
\begin{equation}
a \left( \eta \right) = \sqrt{{3 \, Q^2 \, \omega_4\over{16 \, M}} + \omega_4 \, M \, \eta^2},
\label{fscale}
\end{equation}
Also in this case we have a bouncing universe, with a minimal radius
\begin{equation}
a_{\rm min} = \sqrt{\frac{3 \, Q^2 \, \omega_4}{16 \, M}} \,\,,
\end{equation}
which vanishes in the limit $Q \rightarrow 0 \,$. At late times one
recovers the evolution $a \left( \eta \right) \sim  \eta \sim \tau^{1\over
2}\,$, which is typical of a flat universe dominated by radiation.

\section{Radiation on the brane}~\label{radio}

We now consider the more physically relevant case in which a perfect fluid
with equation of state of radiation is present on the brane. We will
discuss under which conditions the system can describe a universe with a
bounce followed by a radiation dominated phase compatible with
observations (more precisely, with the constraints coming from primordial
nucleosynthesis). In presence of a perfect fluid on the brane,
eq.~(\ref{freedman}) rewrites (see i.e.~\cite{MV} for details)
\begin{equation}
H^2 = - \frac{k}{a^2} + \frac{\omega_4 \, M}{a^4} - \frac{3\,\omega_4^2\,Q^2}{16\,a^6} - \frac{1}{L^2} + \frac{4 \, \pi}{3 \, M_p^2 \, \rho_0} \left( \rho_0 + \rho_{\rm br} \right)^2 \,\,,
\label{frebra}
\end{equation}
where $\rho_0$ denotes the tension of the brane, while the energy density $\rho_{\rm br}$ of the fluid satisfies the standard conservation equation
\begin{equation}
\dot{\rho}_{\rm br} + 3 \, \frac{\dot{a}}{a} \, \left( 1 + w \right) \, \rho_{\rm br} = 0 \,\,,
\end{equation}
where $w$ is the equation of state of the fluid, $w=1/3$ for radiation. 
Notice that the total energy density on the brane $\rho_0+\rho_{\rm br}$
contributes quadratically to $H^2\,$~\cite{BDL}. However, from the mixed
term $\propto \rho_0 \, \rho_{\rm br}$ one recovers at small energies
($\rho_{\rm br} \ll \rho_0$) a leading contribution which is linearly
proportional to the energy density of the fluid~\cite{CGKT,CGS}. The
overall coefficient of the last term of eq.~(\ref{frebra}) has been
normalized so to reproduce the standard coefficient for the term linear in
$\rho_{\rm br}\,$.

The evolution equation~(\ref{frebra}) can be rewritten as
\begin{eqnarray}
H^2 &=& - \frac{k}{a^2} + \frac{\omega_4 \, M}{a^4} - \frac{3\,\omega_4^2\,Q^2}{16\,a^6} + \frac{\Lambda_4}{3} + \frac{8\,\pi}{3\,M_p^2} \left( \rho_{\rm br} + \frac{\rho_{\rm br}^2}{2\,\rho_0} \right) \,\,, \nonumber\\
\Lambda_4 &\equiv& \frac{4\,\pi}{M_p^2} \, \rho_0 - \frac{3}{L^2} \equiv
\Lambda_{\rm br} - \frac{3}{L^2} \,\,.
\label{frebra2}
\end{eqnarray}
The four effective cosmological constant $\Lambda_4$ is obtained as a sum
of the tension of the brane and of the five dimensional cosmological
constant. To have a realistic cosmology, these two quantities have to be
fine-tuned to cancel each other. Even if the cancellation is not perfect,
the contribution from the cosmological constant term to
eq.~(\ref{frebra2}) is negligible at small scale factor $a\,$, so that
this term does not alter the evolution of the universe close to the
bounce. At large $a\,$, we can instead neglect the term proportional to
$Q\,$, as well the last contribution quadratic in $\rho_{\rm br}\,$. Then,
eq.~(\ref{freedman}) effectively describes the rather standard situation
of an universe filled with radiation and cosmological constant. In the
context of brane models, the cosmological evolution for $Q=\rho_{\rm br} =
0$ can be found, for example, in refs.~\cite{PS,MD}.

To discuss in more details the cosmological evolution across and after the
bounce, let us consider eq.~(\ref{frebra2}) at small scale factor. Besides
the constant term $\Lambda_4\,$, also the curvature term  $-k/a^2$ can be
neglected in eq.~(\ref{frebra2}), since it is subdominant also today.
Specifying the energy density on the brane to be in the form of radiation,
$\rho_{\rm br} \equiv \rho_r/a^4\,$, we have
\begin{equation}
H^2 \simeq \left( \frac{8\,\pi\,\rho_r}{3\,M_p^2} + \omega_4 \, M \right) \frac{1}{a^4} - \frac{3\,\omega_4^2\,Q^2}{16} \, \frac{1}{a^6} + \frac{4\,\pi \, \rho_r^2}{3\,M_p^2\,\rho_0} \, \frac{1}{a^8} \,\,. 
\label{frebra3}
\end{equation}
The first contribution is a sum of the effective dark radiation term,
coming from the mass of the five dimensional black hole, and of the
radiation present on the brane. Primordial nucleosynthesis imposes that
the latter term dominates over the former, since radiation is not only
responsible for the expansion of the universe at early times, but also for
the formation of light elements. From the observed abundances of the
latter, one typically concludes~\cite{LSV,OSW} that any nonstandard form
of energy density can contribute at most as an additional neutrino species
at the time of nucleosynthesis (a more constrained bound is found by
combining results from nucleosynthesis with the ones inferred from
anisotropies of the Cosmic Microwave Background~\cite{HMMMP}). Requiring
the dark radiation term to contribute less than an additional neutrino
species gives the constraint
\begin{equation}
\omega_4 \, M \la 1.1 \, \frac{\rho_r}{M_p^2} \,\,.
\label{dis1}
\end{equation}
In the following, we neglect the dark radiation term with respect to the the standard radiation on the brane.

The occurrence of the bounce places another constraint on the parameters.
This is obtained by requiring that the Hubble parameter $H$ given in
eq.~(\ref{frebra3}) vanishes at some finite value of $a\,$, since
otherwise the term proportional to the charge of the black hole $Q$ is
always subdominant with respect to the other two terms. Thus, we must have
\begin{equation}
\omega_4^4 \, Q^4 \ga 4000 \, \frac{\rho_r^3}{M_p^4 \, \rho_0} \,\,.
\label{dis2}
\end{equation}
If this bound is satisfied, the scale factor approximatively amounts to
$a_{\rm min} \simeq 0.15 \, \omega_4 \, Q \, M_p/\rho_r^{1/2}$ at the
bounce (the right hand side of~(\ref{dis2}) has been neglected in this estimate). Since we want a standard (radiation dominated) expansion at the
time of primordial nucleosynthesis, that is when the temperature of the
thermal bath is $\sim 0.2 \,$ MeV, we finally have to require
\begin{equation}
\rho_{\rm br} \left( a_{\rm min} \right) = \frac{\rho_r}{a_{\rm min}^4} =
\frac{\rho_r^3}{0.15^4 \, \omega_4^4 \, Q^4 \, M_p^4} \ga 0.2^4 \, {\rm MeV}^4 \,\,.
\label{dis3}
\end{equation}
By combining the two expressions~(\ref{dis2}) and (\ref{dis3}), we see
that in order for the $\rho_{\rm br}^2$ term in the Hubble equation not to
affect the predictions of primordial nucleosynthesis, the tension of the
brane has to satisfy the lower bound
\begin{equation}
\rho_0^{1/4}  \ga 0.24 \, {\rm MeV} \,.
\end{equation}

This is a less stringent bound than the one coming by imposing that
ordinary gravity is recovered in current gravity experiments,
$\rho_0^{1/4} > {\rm few} \times {\rm TeV} \,$~\cite{CED}.

\section{Fluctuations of brane fields}~\label{pert}

Fluctuations about the regular backgrounds described in the previous
sections are also expected to be regular. Here we discuss the simple case
of fluctuations of a massless minimally coupled scalar field $\phi$ living
on the brane, assuming that it does not contribute significantly to the
cosmological evolution of the background. At the end of the section, we
briefly comment on the more general issue of cosmological perturbations in
this model.  For definiteness, we discuss the case of a flat critical
brane, while the generalization to the other cases is straightforward. The
four dimensional scale factor then evolves according to
eq.~(\ref{fscale}). Here we redefine
\begin{equation}
a \left( \eta \right) = a_{\rm min} \, \sqrt{1+ \gamma^2 \, \eta^2} \;\;,\;\; a_{\rm min} = \sqrt{\frac{3 \, Q^2 \, \omega_4}{16 \, M}}
\;,\; \gamma=\frac{4\,M}{\sqrt{3}\,Q} \,\,.
\label{eqa}
\end{equation}

Consider a mode of the fluctuations $\delta \phi_k$ with a given comoving
momentum $k\,$, and the rescaled mode $v_k \equiv a \, \delta \phi_k \,$.
The latter evolves according to
\begin{equation}
v_k'' + \left( k^2 - \frac{a''}{a} \right) v_k = 0 \;\;,\;\; \frac{a''}{a} = \frac{\gamma^2}{\left( 1 + \gamma^2 \, \eta^2 \right)^2} \,\,.
\label{eqv}
\end{equation}
At early times, the term $a''/a$ is negligible, and the mode behaves as a plane wave in Minkowski space. Since the field $v$ is canonically normalized, we have
\begin{equation}
\delta \phi_k = v_k/a \rightarrow \frac{{\rm e}^{\,-i\,k\,\eta}}{\sqrt{2 \,k} \, a} \;\;,\;\; \eta \rightarrow - \infty \,\,.
\label{plane}
\end{equation}

If $k \gg \gamma\,$, the size of the mode is always much smaller than the 
size of the horizon, and $v$ always behaves as a plane wave. The case $k < \gamma$ is more interesting. As $\eta\,$ increases, the evolution of the universe starts to be important, and the term $a''/a$ cannot anymore be neglected in eq.~(\ref{eqv}). The two terms $a''/a$ and $k$ become equal at the time $\eta_*\,$ given by
\begin{equation}
\gamma \, \eta_* = - \, \sqrt{\frac{\gamma - k}{k}} \,.
\label{etastar}
\end{equation}
In the long wave limit (that is, neglecting $k$ in eq.~(\ref{eqv})) we have then 
\begin{equation}
\delta \phi_k \simeq C_1 + {\tilde C}_2 \, \int^{\,\eta} \frac{d \eta'}{a^2} = C_1 + C_2 \, {\rm arctg } \left( \gamma \, \eta \right)
\;\;,\;\; \vert \eta \vert \ll \vert \eta_i \vert \,\,.
\label{long}
\end{equation}

The coefficients $C_1 \,,\, C_2$ can be estimated by assuming that the
mode evolution is given by eq.~(\ref{plane}) for $\eta < \eta_*\,$ and by
eq.~(\ref{long}) for $\eta > \eta_*\,$, and by matching $\delta \phi_k$
and its derivative at $\eta_*\,$. This gives
\begin{eqnarray}
\vert C_1\vert &\simeq& \frac{1}{\sqrt{2 \, \gamma}\,a_{\rm min}} \, \left[ \frac{\gamma}{k} {\rm arctg}^2 \,
\sqrt{\frac{\gamma - k}{k}} + 2 \, \sqrt{\frac{\gamma - k}{k}} \, {\rm arctg}^2 \,
\sqrt{\frac{\gamma - k}{k}} + 1 \right]^{1/2} \,\,, \nonumber\\
\vert C_2 \vert &\simeq& \frac{1}{\sqrt{2 \, k}\,a_{\rm min}} \,\,.
\label{c1c2}
\end{eqnarray}

As it was expected, the spectrum of these fluctuations is not scale
invariant. For very long wave modes, $k \ll \gamma\,$, we find (up to an
overall irrelevant phase)
\begin{equation}
\delta \phi_k \simeq \frac{1}{\sqrt{2 \, k}\,a_{\rm min}} \, \left[ \frac{\pi}{2} +
{\rm arctg } \left( \gamma \, \eta \right) \right] \,\,.
\label{late}
\end{equation}
We see that the linear theory of the fluctuations breaks down if the background becomes singular at the bounce, $a_{\rm min} \rightarrow 0\,$.

The computation just reported has mainly the aim to enlight some of the
features that can be expected in the more interesting study of the
cosmological perturbations of this system. Although the latter is beyond
the aims of the present work, some considerations are in order. The
simplest approach to this problem is to study the perturbations of a four
dimensional effective theory leading to eq.~(\ref{freedman}). In a four
dimensional context, this can be obtained at the expense of introducing a
massless field $\varphi$ with negative kinetic terms. The absence of a
potential term for the field $\varphi$ guarantees the equation of state
$w_{\varphi} = + 1\,$, so that $\rho_{\varphi} \propto a^{-\,6}\,$, and
eq.~(\ref{freedman}) can be recovered. Cosmological perturbations in this
four dimensional set-up have been studied in~\cite{PEPI}, where it is also
concluded that perturbations remain regular as long as the background is
regular, and where a late time scale depended spectrum is also obtained
(but with different $k$ dependence wih respect to our eq.~(\ref{late})).
Although the results of~\cite{PEPI} are very interesting for the general
study of cosmological perturbations in bouncing models, the presence of a
negative kinetic term opens several problems, and the system should be
considered only as an effective one, as also mentioned in~\cite{PEPI}.
Here, we would like to point out another potential difficulty, related to
the choice of initial conditions. In the standard case, it is customary to
start with an adiabatic (Bunch-Davies~\cite{BUDE}) vacuum, since - for an 
early adiabatic evolution of the background - it minimizes the energy of
the system (for a detailed discussion, see i.e.~\cite{MABR}). With
negative kinetic terms, the adiabatic vacuum maximizes the energy of the
fluctuations, so that the choice of the initial conditions appears more
problematic than in the standard case. An alternative approach is
certainly a direct calculation of the perturbations in the full five
dimensional theory. The formalism for the study of perturbations in brane
models is at the moment under deep investigation, see for
example~\cite{P1,P2,P3}. However, the application of this formalism to the
system discussed in this paper is at present a highly nontrivial issue.

\section{Cardy-Verlinde Formula\label{CVF}}

In this section, we analyse the Cardy-Verlinde formula  \cite{EV,SV} in 
the light of the explicit solutions found in the second section. We
specify to the case of a critical brane. We start with the case of a
closed universe. As discussed in section-2, the brane in this case 
behaves as a closed non-singular universe if $Q$ is less than certain 
critical value. The radius of the brane oscillates between  $a_{\rm min}$
and $a_{\rm max}$ given in eq.~(\ref{minmax}). Note that the bulk geometry
has a  non-singular horizon at $a = a_H$ where $a_H$ obeys
\begin{equation}
1 - {\omega_{4} M\over{a_H^{2}}} + {3 \omega_{4}^2 Q^2
\over {16 a_H^{4}}} + {a_H^2\over{L^2}} =0 \,\,.
\label{horiloc}
\end{equation}
It is easy to check that for a given 
value of $Q$ and $M$, the brane starts expanding at a location beyond
horizon. The distance from the horizon depends critically on the 
$Q/M$ ratio. The radius then increases and, as a result, it comes out 
of the black hole horizon. As time progresses, it then again reaches a 
minimum size 
at a point beyond horizon. 
As we increase this ratio, keeping other 
parameters fixed, $a_{\rm min}$ moves closer to the horizon.

As argued in \cite{EV}, during the cosmological evolution, the  universe
goes from weakly  self-gravitating to strongly self-gravitating region or
otherwise. The  transition occurs  when the Hubble radius $H^{-1}$ is
comparable to the radius $a$ of the  universe. We see from
(\ref{freedman}), this happens when $a$ satisfies
\begin{equation}
a^4 - {\omega_4 M\over 2} a^2 + {3 \omega_4^2 Q^2 \over{32}} = 0.
\label{clocus}
\end{equation}
It is easy to locate at which time $Ha =1$ from our explicit solution 
(\ref{konesoln}). This happens when $\eta$ satisfies
\begin{equation}
\sin~2 \eta + \cos~2\eta  = {1\over{c_1}}.
\end{equation}
During the evolution, the universe passes from 
weak self gravitating phase $(Ha \le 1)$ to strong self-gravitating phase 
$(Ha \ge 1)$. It is easy to check that on the brane universe, 
\begin{eqnarray} 
&&S_B - S_Q \le S ~~~{\rm for} ~~ Ha \le 1, \nonumber\\
&&S_B - S_Q \ge S ~~~{\rm for} ~~ Ha \ge 1.
\label{entb}
\end{eqnarray}
Here, the Bekenstein entropy is $S_B \equiv 2\pi a E/3$, $S_Q \equiv (2\pi
a/3) \phi  Q/2\,$, and $S$ is the Hawking-Bekenstein entropy (\ref{st})
written  in terms of brane Newton constant $G_4 = 2 {\bf G_5}/L$. As
discussed in  \cite{BM}, on the brane, the following relation (similar to
Cardy-Verlinde  formula) holds
\begin{equation}
s = \Big({4 \pi\over n}\Big) \sqrt {\gamma (\rho - {\Phi \tilde \rho\over 
2} - {\gamma\over {a^2}})}.
\label{cv}
\end{equation}
In the above equation, $s, \rho $ and $\tilde \rho$ are the entropy, 
energy and charge densities on the brane that follows from 
(\ref{branest}). Furthermore, 
$\gamma$ represent the Casimir part of the energy density \cite{SV}
and is given by 
\begin{equation}
\gamma = {3 a_H^2\over{8\pi G_4 a^2}},
\label{ce}
\end{equation}
where $a$ is given in (\ref{konesoln}). As $a$ is finite throughout the 
evolution, $\gamma$ is also bounded
\begin{equation}
{9 {\rm Vol}(S^3) a_H^2 \over{ 32 \pi^2 G_4^2 L M(1+c_1)}} \le 
\gamma
\le {9 {\rm Vol}(S^3) a_H^2 \over{ 32 \pi^2 G_4^2 L M (1-c_1)}}.
\end{equation}

We now briefly discuss the Cardy-Verlinde formula for open and flat 
universe. As for $k =0$, universe passes from weak to strong gravitational 
system when $Ha > {\sqrt{2}}$. It can be easily checked that at $Ha =
{\sqrt{2}}$, (\ref{clocus}) holds. From our explicit 
solution (\ref{fscale}), we find that the universe reaches  the 
strong gravity domain when $16 M^2/3 Q^2 > \sqrt 2$. As for $k = -1$, the 
universe transits from weak to strong-gravity region when 
$Ha > \sqrt 3$. This happens, from (\ref{oscale}), when 
\begin{equation}
\cosh~2\eta - \sinh~2\eta = {1\over { c_2}}.
\end{equation}
In both these cases, an equation similar to (\ref{entb}) is satisfied as 
can be found in \cite{DY}. Furthermore, the analogue of (\ref{cv}) is 
given by \cite{DY}
\begin{equation}
s = \Big({4 \pi\over n}\Big) \sqrt {\gamma (\rho - {\Phi \tilde \rho\over
2} - {k\gamma\over {a^2}})}.
\label{cvof}
\end{equation}
The Casimir energy density has the same form as (\ref{ce}) except the 
explicit form of $a_H$ in terms of $M$ and $Q$ is different in this case.
We would now like to point out that, in all our solutions in section-2, 
since universe never goes to zero size, $\gamma$ is always bounded from 
above. In particular, 
\begin{eqnarray}
\gamma &<& {3 a_H^2\over {2 \pi G_4 \omega_4 M (c_2 -1)}}
~~~~{\rm for}~~k = -1,\nonumber\\
       &<& {M a_H^2 \over{2\pi G_4 Q^2 \omega_4}}
~~~~{\rm for}~~k =0.
\end{eqnarray}


\section{Conclusion\label{CONCLUSION}}

In this paper, we have carried out a detailed study of possible
cosmological scenarios for a brane universe in a charged AdS black hole
background. The four dimensional Hubble law for the scale factor of the
brane gets a contribution proportional to the bulk  charge with an 
unconventional negative sign. As a consequence, we found the cosmological 
evolution of the universe to be regular, with a smooth transition between 
a contracting and an eventual expanding phase. We presented exact
cosmological solutions for a (open, flat, and closed) critical brane; the
inclusion of matter on the brane modifies these analytical solutions.
However, in the latter case one can easily recover an evolution
characterized by a bounce followed by a radiation dominated phase
compatible with observations (more precisely, with the constraints coming
from primordial nucleosynthesis). We then discussed the evolution of
fluctuations of fields on the brane, showing that they remain in the
linear regime provided the minimal size of the scale factor is
sufficiently large, which can always be achieved by appropriate choices of
the parameters of the model. The spectrum of the fluctuations is however
not scale invariant. In the last section, we discussed the Cardy-Verlinde
formula in the above set  up. The brane entropy density has a Casimir
part; the energy density of this contribution is always found to be
bounded from above.

\vskip 0.5cm

\noindent{\large\bf Acknowledgment:} 
The work of S.M. is supported  in part by a DFG grant. He also wishes to 
thank Peter Nilles for providing a very pleasant working atmosphere in his
group at Bonn. M.P. acknowledges fruitful discussions with David Lyth,
Lorenzo Sorbo, and Gianmassimo Tasinato. We also thank the referee for
useful comments and suggestions. The work of M.P. is supported by the
European Community's Human Potential  Programme under contracts
HPRN-CT-2000-00131 Quantum Spacetime,  HPRN-CT-2000-00148 Physics Across
the Present Energy Frontier and  HPRN-CT-2000-00152 Supersymmetry and the
Early Universe.



\end{document}